\documentclass[a4paper,11pt]{article}
\usepackage{science}

\newcommand{\br}[1]{\left(#1\right)} 

\newcommand{\half}{\frac{1}{2}}

\newcommand{\ket}[1]{\left|#1 \right\rangle}
\newcommand{\bra}[1]{\left\langle#1 \right|}

\newcommand{\rvec}{\mathbf{r}}

\newcommand{\nvec}{\mathbf{n}}

\newcommand{\Ovec}{\mathbf{0}}
\newcommand{\drvec}{\mathrm{d}^{3N}\rvec}

\newcommand{\Uev}{\hat{U}}

\newcommand{\exnr}[1]{\hat{\psi}^{\dagger}_e(#1)\hat{\psi}_e(#1)}
\newcommand{\indfn}{\mathbf{1}_r}
\newcommand{\indfm}{\mathbf{1}_m}
\newcommand{\seg}[1]{\hat{\sigma}_{\rm 10}^{(#1)}}
\newcommand{\sge}[1]{\hat{\sigma}_{\rm 01}^{(#1)}}
\newcommand{\see}[1]{\hat{\sigma}_{\rm 11}^{(#1)}}

\usepackage{graphicx}

\title{Adiabatic Formation of Rydberg Crystals with Chirped Laser Pulses}

\author{R.~M.~W.~van Bijnen, S.~Smit, K.~A.~H.~van Leeuwen, E.~J.~D.~Vredenbregt, \\ S.~J.~J.~M.~F.~Kokkelmans\\ \textit{Eindhoven University of Technology}, \\ \textit{P.O. Box 513, 5600 MB Eindhoven, The Netherlands}}

\begin{document}
\maketitle

\begin{abstract}
Ultracold atomic gases have been used extensively in recent years to realize textbook examples of condensed matter phenomena. Recently, phase transitions to ordered structures have been predicted for gases of highly excited, ``frozen'' Rydberg atoms. Such Rydberg crystals are a model for dilute metallic solids with tunable lattice parameters, and provide access to a wide variety of fundamental phenomena. We investigate theoretically how such structures can be created in four distinct cold atomic systems, by using tailored laser-excitation in the presence of strong Rydberg-Rydberg interactions. We study in detail the experimental requirements and limitations for these systems, and characterize the basic properties of small crystalline Rydberg structures in one, two and three dimensions.
\end{abstract}

\maketitle


\begin{section}{Introduction}\label{SecIntroduction}

Since the development of techniques to cool and trap atomic gases with laser radiation in the 1990's, it has been possible to realize and study many-body phenomena in dilute, ultracold gases. Some of these phenomena are straight out of solid-state textbooks, such as the Mott-Hubbard transition \cite{Jaksch98,Greiner02}: the transition from a ``conducting'' state to an ``insulator'' state. In optical lattices, which are artificial crystals of light created by standing-wave laser beams, bosonic atoms could be trapped with a filling factor of unity, and a superfluid to Mott-insulator transition could be observed.

One serious shortcoming of such cold atom experiments with regard to the comparison with condensed matter physics is that neutral ground-state atoms interact via short-range Van der Waals interactions, which are of course much weaker than the Coulomb interactions between electrons in solids. A much closer realization of a strongly-coupled condensed matter system would arise if the atoms could be separated into ions arranged on a lattice, and surrounding electrons that provide the conductance phenomena.

The question then arises how this could be achieved. Ultracold gases can be made strongly-interacting by using Feshbach resonances~\cite{Feshbach58,Tiesinga92}, however, it does not mean that the system is automatically strongly-coupled~\cite{Killian07}. These gases are typically very dilute, such that the Van der Waals range $r_0=(m C_{6}/\hbar^{2}) ^{1/4}/2$ of the ground-state atoms, where $m$ is the atomic mass and $C_6$ the Van der Waals coefficient, is much smaller than the average distance between the particles $n^{-1/3}$, with $n$ the density. The coupling parameter $\Gamma$ that indicates the transition between weakly coupled and strongly-coupled systems, is given by the ratio of interaction energy $E_{\rm int}$ and kinetic energy $E_{\rm kin}$. Even for a unitary Fermi gas~\cite{Giorgini08}, which is by definition strongly interacting, the coupling parameter is smaller than 1: $\Gamma=|E_{\rm int}/E_F|\approx 0.6$, with $E_{\rm kin}=E_F$ the Fermi energy.

One way to study strongly interacting and strongly coupled atomic systems is by using ultracold Rydberg atoms~\cite{Gallagher88}. Rydberg exitations into electronic $s-orbital$ states interact, similar as the ground-state atoms, via Van der Waals interactions. However, the Van der Waals coefficient scales very rapidly with the principal quantum number as $C_6 \sim n^{11}$, and therefore it is possible to reach the regime $r_{0} > n^{-1/3}$.
Robicheaux and Hern\'andez predicted already in 2005 many-body correlations which can rise from the Rydberg blockade effect (also referred to as dipole blockade effect) in a disordered gas~\cite{Robicheaux05}, hinting at the possibility of ordered structures. Later, Weimer {\it et al.}~showed that a phase transition could occur from a strongly interacting Rydberg gas to a crystalline phase~\cite{Weimer08}, for positive detuning from the ground-state to Rydberg transition. Since then several proposals have been made
to create correlated systems that involve Rydberg atoms: for instance by means of dynamical crystallization of Rydberg atoms starting from ground-state atoms~\cite{Pohl10,Schachenmayer10,Stanojovic10}, or by weakly dressing ground-state atoms to form super-solid droplet crystals~\cite{Pupillo10,Henkel10,Cinti10}.

In this paper, we investigate a very practical scheme which was first proposed by Pohl, Demler and Lukin~\cite{Pohl10}, which allows for a laser-assisted, adiabatic self-assembly of a Rydberg atom lattice. We investigate the robustness of a tailored excitation scheme to build up a correlated crystal state from a disordered atomic gas. Although this paper has a theoretical nature, our investigations are based on the present ultracold atom set-up in our group in Eindhoven. The experiment contains also an accelerator infrastructure (see Fig.\ref{fi:exp_setup}), which provides us with a tool to detect correlated Rydberg crystal structures.
\begin{figure}
\begin{center}
\includegraphics[width=\columnwidth]{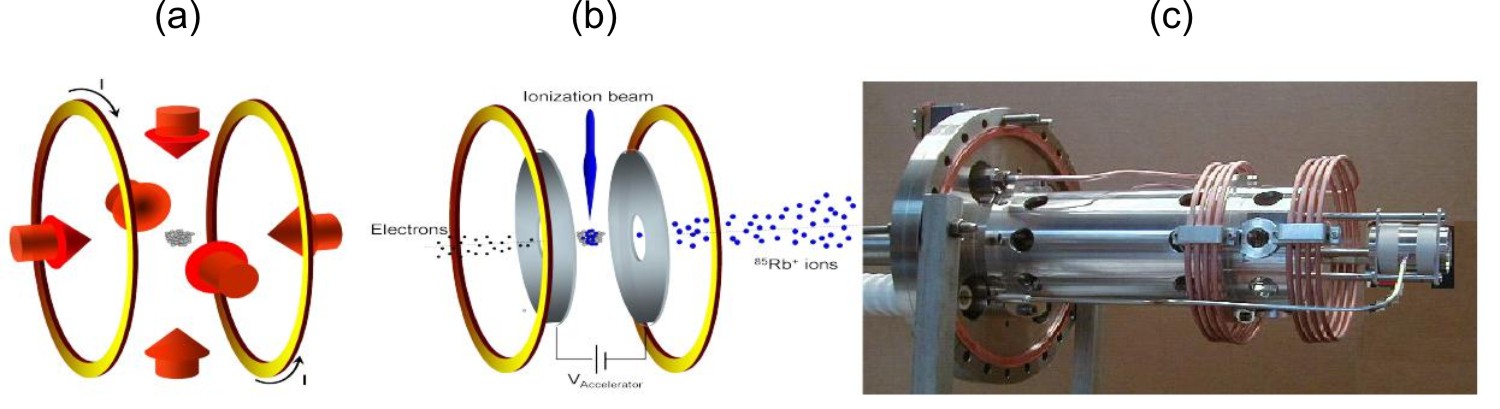}
\caption{Schematic overview of our experimental set-up, on which the model parameters are based. (a) Atoms are cooled and trapped in a Magneto-optical trap. (b) Then an ionization beam is utilized to create an ultracold plasma. Due to the accelerator, the electrons and ions can be used to create bright beams of electrons or ions. The ultracold particle accelerator ((pictured in (c)) provides us with a tool to spatially resolve the Rydberg lattices. }\label{fi:exp_setup}
\end{center}
\end{figure}
The set-up contains a magneto-optical trap for rubidium atoms with appropriate 780 nm trapping lasers. We routinely trap $\approx 10^8$ $^{85}$Rb atoms at densities up to $10^{17}$ atoms/m$^3$ and with temperatures below 1 mK. Both pulsed (10 ns YAG-pumped dye) and continuous-wave, single-frequency (300 mW solid-state) 480 nm lasers are present for Rydberg excitation~\cite{Taban10}. The atom trap is built inside an accelerator, followed by a beam transport system that allows one to create magnified images of both ions and electrons onto a microchannel plate and phosphor screen assembly that is observed with a cooled CCD camera. The acceleration field can be switched on to 10kV/cm on sub nanosecond time scales. This set-up has been used to study electron and ion beams generated from the laser-cooled atoms by near-threshold photo-ionization and by field-ionization of Rydberg atoms. Both ions and electrons can be detected with a multichannel plate as a measure for the total number of Rydberg atoms produced. By directly ionizing the sample, and thus creating an ultracold plasma, followed by a rapid extraction of charged particles, we demonstrated that this technique can be used to produce cold bright beams of ions and electrons, with temperatures in the beam of the ions of T=1mK and for the electrons of T=10K~\cite{Taban10, Reijnders09}.

This paper is organized as follows: In Section \ref{Systemdescription} we give a theoretical desciption of a cold atomic gas in the frozen gas limit, and the corresponding Hamiltonian that describes it. In Section \ref{SecCreatingCrystals} we argue how Rydberg crystals can be created by adiabatically chirping  the laser frequency. In Section \ref{SecPhysicalSystems} we look in detail into four distinct cold atomic systems which we consider for creating Rydberg lattices. In Section \ref{SecPerformingChirp} we present our results for crystal formation using tailored excitation pulses for one-dimensional systems. In Section \ref{SecHigherDimensionalSystems} we investigate a few basic crystal properties of two- and three-dimensional systems, and finally we conclude in Section \ref{SecConclusions}.

\end{section} 

\begin{section}{System description}\label{Systemdescription}
Throughout this paper we will be working in the frozen gas limit, where the excitation dynamics happen on a time scale so fast that we can neglect all motion of the atoms \cite{Pohl10, WeidemullerBook}, i.e.,~ignore their kinetic energy.

We will investigate four different systems: (i) Atoms in a deep optical lattice. (ii) Condensed atoms in a BEC, where the Thomas-Fermi approximation can be made. (iii) Thermal atoms in a dipole trap, where temperatures can be in the $\mu$K regime. And finally, (iv) Thermal atoms in Magneto-Optical Trap (MOT) conditions, with temperatures typically in the sub-mK regime.

Under all of the above conditions, the frozen gas approximation is satisfied. Even for the highest temperatures in the MOT, it takes tens of $\mu$s for atoms to move significantly on the crystal scale of $~5-10 \mu m$ \cite{Pohl10, WeidemullerBook}. The Doppler limit, which gives rise to the temperature of the atoms in the MOT, can be expressed as a velocity that is about 0.14 m/s for rubidium. With this velocity, at the time scales of interest of $~1 - 10 \mu$s, we can safely assume that the interaction energy between Rydberg atoms is constant. Nevertheless, a dephasing mechanism based on Doppler broadening due to the thermal motion of the atoms could potentially frustrate lattice formation. In the following sections, we show how crystal creation relies on adiabatic changing of system parameters. In  \ref{dephasing} we verify, in a two-particle picture, that Doppler broadening does not affect such dynamics.

In general, the systems that we describe consist of $N$ atoms in a laser field. We assume an effective laser coupling between the ground state and a particular Rydberg state, such that we can describe each atom as a two level system. The $N$-body excitation state can then be represented by basis states of the form $\ket{\nvec}$, where $\nvec$ is a $N$-dimensional vector with elements $n_i = \{0,1\}$, indicating whether the $i$-th atom is in the ground (0) or excited (1) state.
First, we consider the special situation in which the $N$ atoms are localised at fixed positions. We will shortly show how this picture can be applied to more general systems. The positions of the $N$ atoms are denoted with the $3 \times N$ vector $\rvec$, where $\rvec_i \in \mathbf{R}^3$ is the position of the $i$-th atom.
In the rotating frame, the Hamiltonian governing the dynamics of the excitation state vectors $\ket{\nvec}$ of the atoms is given by \cite{Robicheaux05, Pohl10}
\begin{equation}\label{EqH}
\hat{H}_{exc}(\rvec) = -\sum_{i=1}^N \hbar \Delta \see{i} + \half \sum_{j=1}^N \hbar \Omega \br{\seg{j} + \sge{j}} + \mathop{\sum_{i=1}^N}_{j > i}  \frac{C_6}{|\rvec_i - \rvec_j|^6} \see{j}\see{i},
\end{equation}
where $\see{i} = \ket{1_i}\bra{1_i}$ counts whether the $i$-th atom is excited, and $\seg{i} = \ket{1_i}\bra{0_i}$ and $\sge{i} = \ket{0_i}\bra{1_i}$ change the state of the $i$-th atom. The first two terms in Eq. (\ref{EqH}) describe the atom-laser interaction, with $\Delta$ the laser detuning and $\Omega$ the effective Rabi frequency of the system. The third term equals the Van der Waals interaction between atoms in the Rydberg state.

In practice there are often many individual atoms spaced so closely together, that we can safely say that within a certain group of atoms there will never be more than one excitation due to the Rydberg blockade. Moreover, we cannot (e.g.~because of detection resolution), or need not, distinguish which particular atom carries an excitation. Such groups of atoms can be said to coherently share an excitation, and are commonly called superatoms \cite{Heidemann07}. We can partition the space into $M$ volumes $V_k$, $k = 1 .. M$, each of which contains a group of $M_k$ atoms which together represent the $k$-th superatom. The Hamiltonian (\ref{EqH}) can be equivalently applied to describe superatoms instead of single atoms, where the positions $\rvec_i$ are taken to be weighted averages of atomic positions within a superatom, and the Rabi frequency $\Omega$ is replaced by a collective Rabi frequency $\Omega_k$ for the $k$-th superatom \cite{Robicheaux05,Heidemann07,Diecke54,Vuletic06,Olmos10PRA}:
\begin{equation}\label{SuperatomOmega}
\Omega \rightarrow \Omega_k=\sqrt{M_k} \Omega.
\end{equation}
This collective Rabi frequency reflects the fact that there are multiple atoms in a superatom, and that the superatom as a whole is more susceptible to the laser light. This latter fact becomes important later on, once we consider dense, inhomogeneous systems. Moreover, in Sec. \ref{SecPerformingChirp} we shall perform simulations of the $N$-body system and superatoms will prove to be a useful means for reducing the size of the Hilbert space under consideration.

Having introduced the Hamiltonian for a collection of particles localised at fixed positions, it is now time to turn our attention to more general systems such as Bose-Einstein condensates with fully coherent wavefunctions, or statistical mixtures found in thermal clouds. To this end, we expand the Hilbert space $\mathcal{H}_e$ of excitation configurations $\ket{\nvec}$, with the Hilbert space of $N$-body spatial wave functions $\mathcal{H}_r$, to form the total product Hilbert space $\mathcal{H}_{total} = \mathcal{H}_r \otimes \mathcal{H}_e$.

A convenient basis for this statespace is given by states
\begin{equation}\label{EqBasis}
\ket{\rvec} \otimes \ket{\nvec},
\end{equation}
where $\ket{\rvec}$ is the state with all $N$ particles localised at positions specified by the $3N$-dimensional vector $\rvec$. The Hamiltonian operator (\ref{EqH}) is straightforwardly extended to the total Hilbert space:
\begin{equation}\label{EqHtotal}
\hat{H} = \int \drvec \ket{\rvec} \hat{H}_{exc}(\rvec) \bra{\rvec}.
\end{equation}
It is in this final step where the frozen gas approximation becomes apparent: in absence of kinetic energy, the Hamiltonian $\hat{H}$ is diagonal in the spatial components.

Using the basis states (\ref{EqBasis}), and assuming that all particles start in the ground state $\ket{\nvec} = \ket{\Ovec}$, we can write the general state of the system at $t=0$ as
\begin{equation}\label{EqPsi0}
\ket{\Psi(0)} = \int \drvec \ c_0(\rvec) \ \ket{\rvec}\otimes \ket{\Ovec}.
\end{equation}
When $t>0$, this state starts to evolve under the action of the Hamiltonian (\ref{EqHtotal}), written down formally with the time evolution operator $\Uev(t)$:
\begin{equation}\label{EqPsit}
\ket{\Psi(t)} = \Uev(t) \ket{\Psi(0)} = \int \drvec \ c_0(\rvec) \ \Uev(t) \Big(\ket{\rvec}\otimes \ket{\Ovec}\Big).
\end{equation}
The frozen gas Hamiltonian (\ref{EqHtotal}) however cannot make transitions between states with different spatial parts $\ket{\rvec}, \ket{\rvec'}$, with $\rvec \neq \rvec'$. Therefore, the time evolution operator of Eq.~(\ref{EqPsit}) will only affect the excitation part $\ket{\nvec}$ of the many-particle state $\ket{\rvec}\otimes\ket{\nvec}$, and not the spatial part $\ket{\rvec}$, so that we can write
\begin{equation}
\Uev(t) \Big(\ket{\rvec}\otimes \ket{\Ovec}\Big) = \ket{\rvec} \otimes \Uev^{exc}_\rvec(t)\ket{\Ovec}.
\end{equation}
Here, the time evolution operator $\Uev^{exc}_\rvec(t)$ governs the excitation state of all the particles under the action of the Hamiltonian (\ref{EqH}), given the fact that they are pinned at definite locations $\rvec$. The general time-dependent state in Eq.~(\ref{EqPsit}) can then be written as
\begin{equation}\label{EqPsiSuper}
\ket{\Psi(t)} = \int \drvec \ c_0(\rvec) \ket{\rvec}\otimes \ket{\chi_\rvec(t)},
\end{equation}
where we have defined excitation state vectors
\begin{equation}\label{EqChir}
\ket{\chi_\rvec(t)} := \Uev^{exc}_\rvec(t) \ket{\Ovec} = \sum_{n=0}^{2^N-1} c_{n, \rvec}(t) \ket{\nvec},
\end{equation}
that describe the excitation state of a given atomic configuration $\ket{\rvec}$. The frozen gas Hamiltonian (\ref{EqHtotal}) only affects the excitation part $\ket{\chi_\rvec(t)}$, which thus carries the time dependence whereas the spatial part $\ket{\rvec}$ remains ``frozen''. It should be noted that $c_0(\rvec)$ and $c_{n, \rvec}(t)$ thus satisfy separate normalisation conditions given by
\begin{equation}
\int \drvec \ |c_0(\rvec)|^2 = 1, \hspace{1cm} \sum_{n=0}^{2^N-1} |c_{n, \rvec}(t)|^2 = 1.
\end{equation}
From the above discussion, we can conclude that the time evolution of a system in the state (\ref{EqPsi0}) can be broken down into the time evolution of the individual, position basis components $\ket{\rvec}$ that make up the total state. Each such component evolves completely independent from the others as a result of the diagonality of the Hamiltonian (\ref{EqHtotal}). Expectation values of observables of interest, $\hat{O}$, that are independent of momentum, such as the excitation density, are computed for each of the individual time evolved configurations. The results are combined in an incoherent sum, in which the coefficients $c_0(\rvec)$ only appear in terms of their modulus squared:
\begin{equation}\label{EqObs}
\langle \hat{O} \rangle = \int \drvec |c_0(\rvec)|^2\bra{\rvec \otimes \chi_\rvec(t)} \hat{O} \ket{\rvec \otimes \chi_\rvec(t)}.
\end{equation}
Similarly, expectation values of observables in statistical mixtures, such as thermal clouds in a trap, are also computed using incoherent sums over individual time-evolved configurations. Here, the coefficients $|c_0(\rvec)|^2$ in Eq. (\ref{EqObs}) are replaced by the statistical probabilities of finding a particular atomic configuration $\rvec$. In the remainder of this paper we shall therefore 
employ Eq. (\ref{EqObs}) for both coherent states as well as thermal clouds, with the understanding that the $|c_0(\rvec)|^2$ coefficients are statistical probabilities in the latter case.
\end{section} 

\begin{section}{Adiabatically creating Rydberg crystals}\label{SecCreatingCrystals}
In the previous Section we have argued how the time evolution of the many-particle state can be broken down into the time evolution of the individual,  position-basis states $\ket\rvec\otimes\ket{\chi_\rvec(t)}$ that make up the total state $\ket{\Psi}$ of Eq. (\ref{EqPsiSuper}).
To see how a crystal state can adiabatically evolve in the frozen gas limit, we now single out a particular component configuration $\ket{\rvec}$ of the atoms, and analyse the time evolution of the corresponding excitation state $\ket{\chi_\rvec(t)}$.
Since we restrict ourselves to one particular configuration $\ket{\rvec}$, we shall omit the spatial ket from the product state notation.

In the limit of zero laser intensity, $\Omega \to 0$, the uncoupled states $\ket{\nvec}$ are eigenstates of the Hamiltonian (\ref{EqH}). The energy $E_n$ of each state $\ket{\nvec}$ assumes a simple linear dependence on the detuning:
\begin{equation}\label{EqEn}
E_n = -m \Delta + E_{int},
\end{equation}
where $m = \sum_i n_i$ is the total number of excited atoms in the state $\ket{\nvec}$, and $E_{int}$ is the total interaction energy between the excited atoms in this particular configuration. Clearly, the energies of all states with the same number of excited atoms have the same slope as function of the detuning $\Delta$. However, depending on the specific location of the $m$ excitations, each state has a different offset, $E_{int}$, at $\Delta = 0$ due to the interaction energies.

\begin{figure}[h]
\begin{center}
\includegraphics[width=11cm]{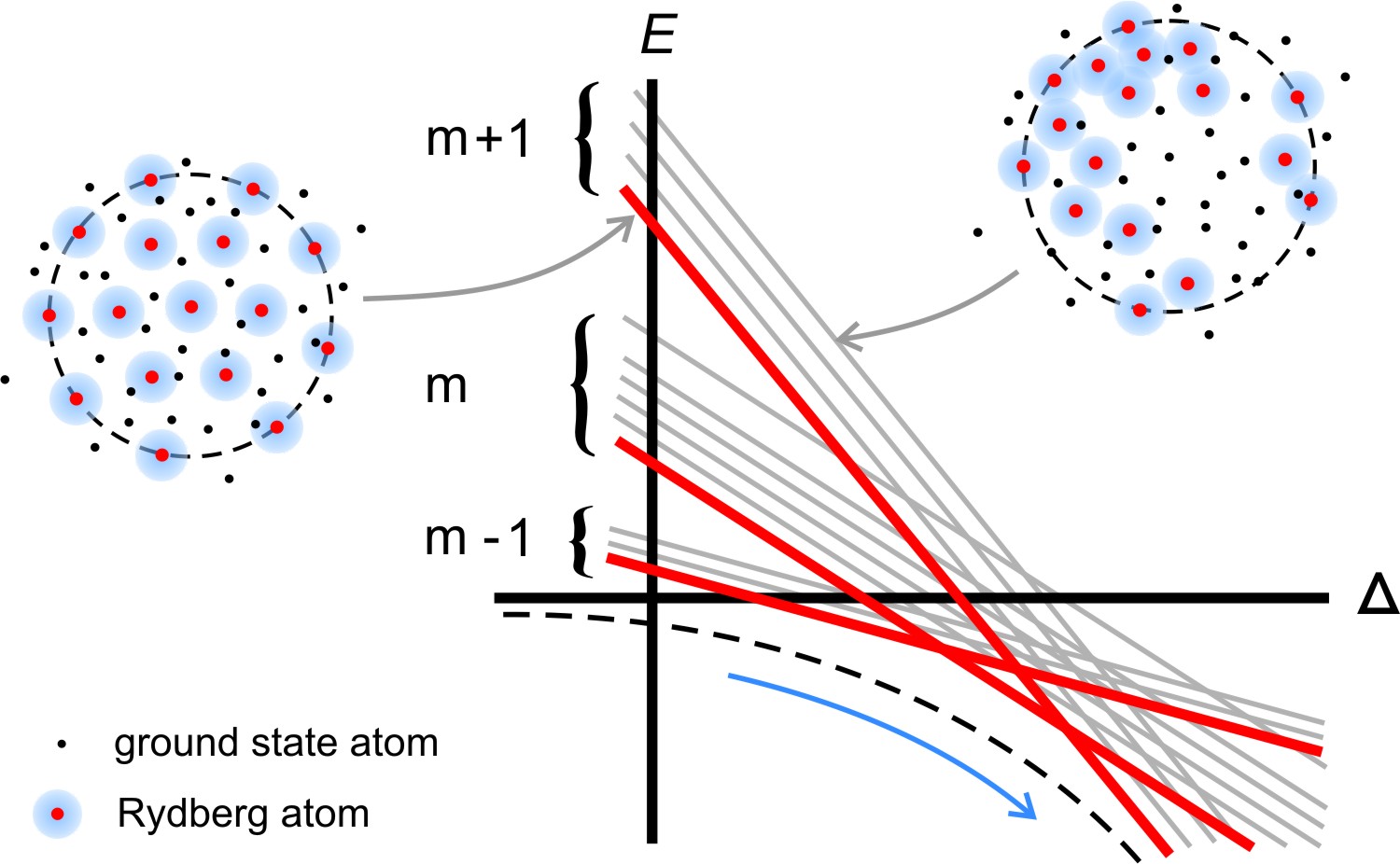}
\caption{\textit{(Schematic) Energy of the basis states $\ket{\rvec}\otimes\ket{\nvec}$ (thin, gray lines), as a function of the detuning and for a given, fixed, choice of the frozen atom positions $\rvec$. All energies depend linearly on the detuning with a slope given by the number of excitations $m$, and the offset at $\Delta=0$ is determined by the Van der Waals interaction energy between the Rydberg atoms. For each value of $m$ there exists a specific configuration with the lowest energy (thick, red line), in which the Rydberg excitations are typically regularly ordered (schematically pictured configuration on the left). Less ordered states have higher energies (example configuration on the right). Note that the atomic positions do not change, merely the excitation localisation is different between the two configurations. For non-zero laser coupling $\Omega > 0$, the ground state (dashed line) becomes a superposition of basis states and separates in energy, which can be adiabatically followd in the direction of the arrow when slowly changing the detuning.}} \label{FigAdiabaticLatticeCreation}
\end{center}
\end{figure}

As such, all states with a specific excitation number $m$ form a manifold in the $E-\Delta$ plane, as illustrated  in Fig.~\ref{FigAdiabaticLatticeCreation} by sets (a ``manifold'') of gray lines (individual states) with the same slope. Now, there are two crucial points to be noted.

Firstly, the lowest lying state within each manifold is typically \textit{an ordered, crystalline state}, indicated with a solid, red line in Fig. \ref{FigAdiabaticLatticeCreation}. That is, the excitations will tend to be found at regular spacings (in Sec. \ref{SecHigherDimensionalSystems} we discuss the crystal structure in more detail). The states with the same excitation number, but a less ordered distribution of excitations will typically have a much higher energy. This is indicated schematically with two example configurations in Fig. \ref{FigAdiabaticLatticeCreation}. For each value of the detuning, there will be a ground state
with one specific excitation number, and one specific configuration of excitations, which has the lowest energy of all possible states. For $\Omega \to 0$, the ground state will always be an ordered, crystalline state.

The second important point concerns what happens when $\Omega > 0$, and regards the possibility to adiabatically create a crystal state. As the laser intensity $\Omega$ is increased from zero, it couples the different states $\ket{\nvec}$ through the second term in the Hamiltonian (\ref{EqH}). Crossings between states from adjacent manifolds in Fig. \ref{FigAdiabaticLatticeCreation} now turn into anticrossings. In particular, the ground state now becomes a superposition of states $\ket{\nvec}$, and separates in energy from the others, as indicated by a black, dashed line in Fig. \ref{FigAdiabaticLatticeCreation}, lying below all others.

A scheme to generate a crystal state would then be as follows.
We start the system with a large negative detuning, $\Omega = 0$, and all atoms in their atomic ground state, such that we are also in the ground state of the many particle system. When the laser coupling is then turned on, and subsequently the detuning is raised slowly enough, the many particle state will adiabatically follow the ground state~\cite{Pohl10}, as indicated by the blue arrow in Fig. \ref{FigAdiabaticLatticeCreation}. This process can continue until some final detuning is reached, at which point the laser coupling $\Omega$ is switched off again. If this last step is also performed adiabatically, the system will end up in the $\Omega=0$ ground state, which is one of the previously discussed ordered crystal states, with a given number of Rydberg excitations each of which is localised on exactly one of the frozen atoms, and regularly spaced, forming a crystal.

The preceding discussion has focused on one particular configuration of atoms $\ket{\rvec}$. However, the system state is typically built up from many different configurations, either as a superposition of the form (\ref{EqPsi0}) in the case of a BEC, or a statistical mixture in the case of a thermal ensemble. As explained in the previous section, expectation values of observables for such systems are formed by incoherent sums of the results computed for such individual configurations. Typically, crystal structures are retained in this summation procedure.

In the next section we will outline experimental conditions under which adiabatic crystal preparation is theoretically possible, followed by Sec. \ref{SecPerformingChirp} where we will perform simulations of such experiments.

\end{section} 

\begin{section}{Physical systems}\label{SecPhysicalSystems}
In this section we discuss in more detail the different experimental systems where Rydberg crystal creation is possible. We strict ourselves to a (quasi) 1D geometry for simplicity as well as for computational feasibility, however, it should be noted that the principle of crystal creation is straightforwardly extended to higher dimensions. We consider $^{85}$Rb atoms initially in the ground state 5s~$^2S_{1/2}$, which we can excite to the $n=65$ Rydberg state $n$s~$^2S_{1/2}$ by using a two-step laser excitation scheme. The first step involves an off-resonant 5s-5p 780 nm transition with Rabi frequency $\Omega_1$ and detuning $\hbar \delta$, and the second step involves a 5p-ns 480 nm transition with Rabi frequency $\Omega_2$. The two-photon Rabi frequency is then given by $\Omega=\Omega_1 \Omega_2/2\delta$. For the $n=65$ Rydberg state, the Van der Waals coefficient is given by $C_6=2.4\ 10^{-58}$~Jm$^6$ \cite{Reinhard07}. We also consider a fixed 1D excitation volume, of length a=45 $\mu$m, which will be defined by the excitation laser profiles and / or the spatial dimensions of the atomic system.

The four main types of systems in ultracold atom experiments are (i) optical lattices, (ii) Bose-Einstein condensates, (iii) thermal clouds in an optical dipole trap (high density and low temperature), and (iv) thermal clouds in a MOT (low density and relatively high temperature). For each of these systems we will show under which conditions Rydberg crystal creation should be possible, and we provide some typical numbers as well as investigate possible drawbacks and advantages.

\textit{(i) Optical lattices:}  The conceptually simplest system is formed by ultracold atoms trapped in a deep optical lattice. Here we consider a slightly idealised system as an example from which we can understand the more complicated systems. We assume that within the excitation volume there is a fixed and uniform number of atoms $N_A$ per site, and that the excitation volume is sharply defined. The lattice is assumed so deep that the atoms are in the Mott insulator phase and there is no hopping, such that the atoms are trivially in the frozen gas limit. The system is then represented by a regularly spaced chain of superatoms, each representing $N_A$ individual atoms. Typically, $N_A$ is simply equal to $1$, and lattice spacings of $~1$ $\mu m$ are commonly realised \cite{Bloch05}, such that there in the order of $40 - 50$ atoms within the excitation volume, all experiencing the same coupling to the laser field.

The crystal states in turn are formed by regularly spaced excitations of the atoms. The critical detunings $\Delta_m^{m+1}$ for which the lowest energy excitation configuration changes from $m$ to $m+1$ excitations, can be estimated as \cite{Pohl10}
\begin{equation}\label{EqDeltan}
\hbar \Delta_m^{m+1} = \frac{C_6}{a^6} \br{m^7 - (m-1)^7}.
\end{equation}
Here we have assumed that the excitations are regularly spaced, and ignored next-nearest neighbour interaction energies. Thus, to create a crystal of $m$ excitations, the final value of the detuning at the end of a chirp should lie between $\Delta_{m-1}^m$ and $\Delta_m^{m+1}$.

Finally, we like to mention that crystal creation in optical lattices in a 1D geometry has already been extensively investigated by Schachenmayer et al. in reference \cite{Schachenmayer10}, finding interesting effects due to descrepancies between lattice and excitation spacing. Similarly, quasi 1D systems in a ring geometry have been thoroughly studied by Olmos et al. \cite{Olmos09,Olmos10NJP,Olmos10PRA} who, for instance, make use of symmetries to reduce computational costs.

\textit{(ii) Bose Einstein Condensates:} The next type of system we consider is a Bose-Einstein condensate (BEC) in the Thomas-Fermi limit \cite{Stringari}. The main difference with the optical lattice is a much higher number of particles in the excitation volume, as well as a non-uniform continuous density distribution. As we saw in the superatom picture described in Sec.~\ref{Systemdescription}, this affects the way the system responds to laser light. In the Thomas-Fermi (TF) limit, the density profile $n(\rvec)$ assumes  a parabolic form as a function of position $z$ along the tube: $n(z) = n_0 \br{1 - \br{\frac{z}{R_z}}^2}$, where $R_z$ is the TF-radius of the condensate and $n_0$ is the central density. The excitation volume can be created by a wide 780 nm beam illuminating the entire condensate, and a narrowly focused 480 nm laser passing through the cloud, requiring $R_z = 22.5\mu m$ to obtain a $45 \mu$ m long tube. The diameter of this tube will be set to 4 $\mu$m, which is reasonable regarding the diffraction limit for this wavelength.

 Typically, $n_0 \approx 10^{14}$ cm$^{-3}$ \cite{Heidemann07, Heidemann08}, leading to $O(10^4)$ particles in the excitation volume and a condensate well within the Thomas-Fermi limit.

Since the density of the system is not uniform, the many-particle position configurations $\ket{\rvec}$ of Eq.~(\ref{EqPsiSuper}) also have a non-uniform probability distribution $c_0(\rvec)$ associated with them. In particular, the probability distribution $P$ of the position $r_L$ of the leftmost particle is equal to
\begin{equation}\label{EqProbrL}
P(r_L \leq r) = F(r_L)^N,
\end{equation}
where $F(r)$ is the cumulative distribution function of the single particle probability distribution. The probability density for the rightmost particle is found by mirroring Eq.~(\ref{EqProbrL}) around $r=0$.
These probability distributions are important, as the outermost particles of a particular configuration $\ket{\rvec}$ are certain to carry an excitation in the corresponding ground state. As a result, the crystal size $a$ used in calculating the critical detunings $\Delta_m^{m+1}$ in Eq. (\ref{EqDeltan}) will vary between different configurations $\ket{\rvec}$, and the critical detunings will likewise assume a probability distribution for their values. Fig.~\ref{FigDminDmax}(a) shows the probability distribution of critical detunings $\Delta_4^5, \Delta_5^6$, and $\Delta_6^7$, relevant for creating a $5$ or $6$ excitation crystal, calculated with $3.6 \cdot 10^4$ particles in the excitation volume.
\begin{figure}[h]
\begin{center}
\includegraphics[width=10cm]{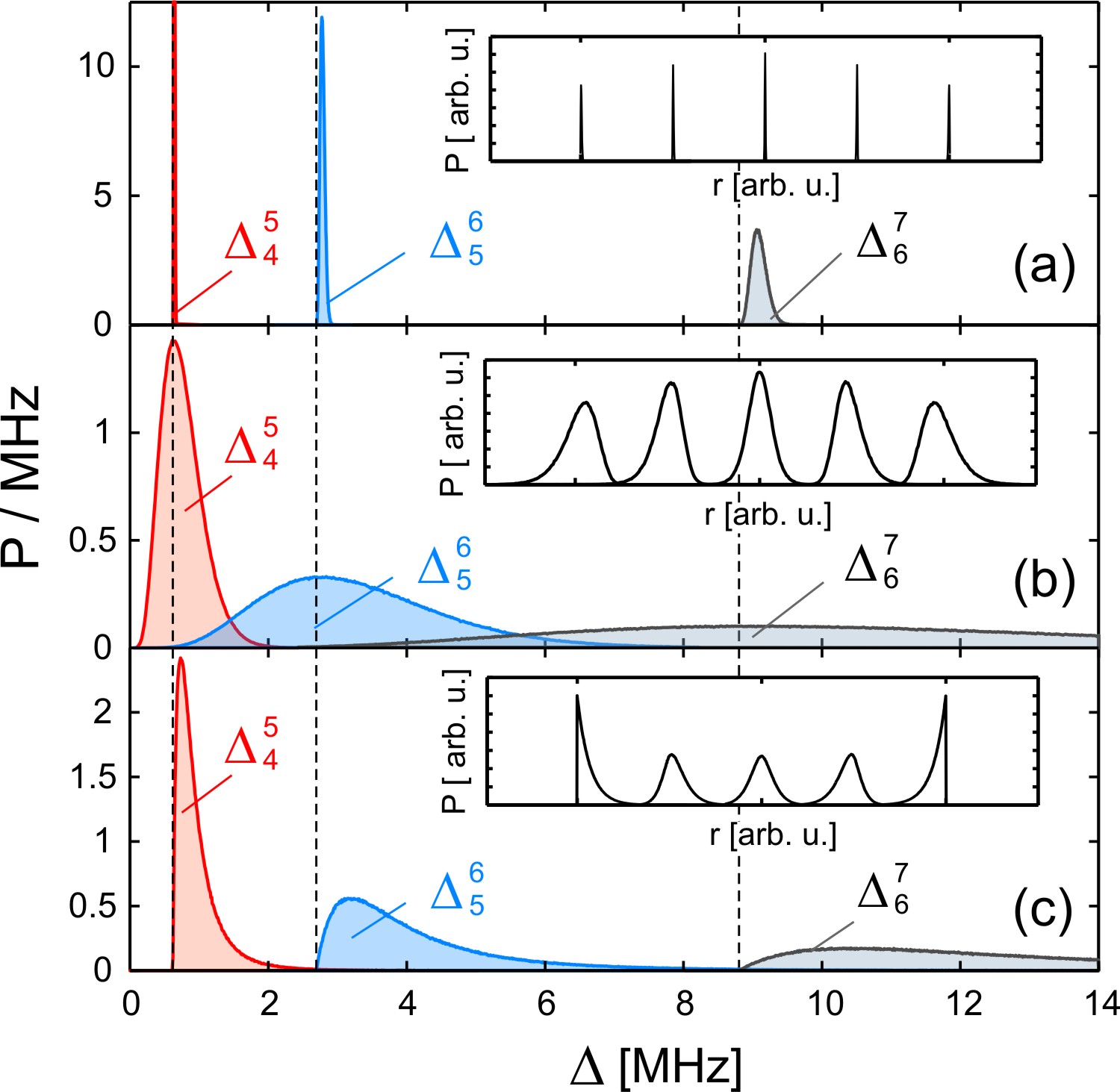}
\caption{\textit{Target detuning probability densities for (a) a BEC in the Thomas-Fermi limit, (b) a thermal cloud in an optical dipole trap, and (c) a thermal cloud in a magneto-optical trap. The critical values for the optical lattice, as given by Eq. (\ref{EqDeltan}) are superimposed as vertical, dashed lines. Insets: probability distribution for the locations of excitations in the $5$-excitation ground state.}} \label{FigDminDmax}
\end{center}
\end{figure}
As opposed to the optical lattice case, the distributions now have a certain width, for instance the $\Delta_6^7$ value has a spread of about $0.5$ MHz about its mean value. To create a $6$ excitation crystal with high fidelity one therefore has to have a final detuning at least $0.5$ MHz away from the mean value of $\Delta_6^7$. The peaks are well enough separated that it is at least always possible to find a suitable detuning at which we are certain to have a definite excitation number in the ground state. As we will shortly see in the case of thermal clouds, this need not always be the case. The inset of Fig.~\ref{FigDminDmax}(a) shows the probability distribution for the locations of the excitations in the $5$-excitation ground state. Due to the probability distribution of the outermost particles, the peaks are not infinitely sharp but have a finite width.

\textit{(iii - iv) Thermal clouds in optical dipole traps and MOTs:} Thermal clouds with Gaussian density distributions form the starting point for many experiments in ultracold atoms. There is a great variety in sizes and density, so we will treat two specific cases, clouds confined in an optical dipole trap (ODT) and clouds contained in a magneto-optical traps (MOT).

Clouds confined in an ODT are much smaller in size than those in a MOT, and they reach much higher densities. Typical peak densities $n_0$ in ODTs range between $10^{12} - 10^{15} \mathrm{cm}^{-3}$ \cite{Heidemann07, Barrett}.

We consider a cylindrically symmetric density of the form
\begin{equation}
n(r, z) = \frac{N}{(2\pi) \kappa^2 \sigma_z^3} \exp(-r^2 / 2 \kappa^2\sigma_z^2 - z^2 / 2 \sigma_z^2),
\end{equation}
where $\kappa$ is the aspect ratio of the cloud, $N$ the total number of atoms in the cloud, and $\sigma_z$ the width of the distribution in the $z$-direction. To create an excitation volume of $45 \mu$m long in the ODT, we can proceed in the same fashion as was the case of a BEC, by using a defocused 780 nm beam illuminating the entire sample, and a narrowly focused 480 nm beam aimed through the center of the cloud along the $z$-axis. For typical numbers $N = 4\cdot 10^5$ atoms, an aspect ratio of $\kappa=10$ and a width of $\sigma_z = 7 \mu$m, we get $2000$ particles in a tube of $4 \mu$m in  diameter. These values are precisely such that the expectation values of the positions of the outermost particles in the tube are spaced $45 \mu$m apart.

As was the case for the BEC, the nonuniform probability distribution for the particle density causes the critical detunings of Eq. (\ref{EqDeltan}) to assume a probability distribution as well. Fig.~\ref{FigDminDmax}(b) shows the probability density for the critical detunings $\Delta_4^5, \Delta_5^6$, and $\Delta_6^7$ relevant for creating a $5$ or $6$ excitation crystal in our excitation volume in the optical dipole trap. Apparently, in this case the probability densities overlap, which means that there exist no values of $\Delta$ in which the ground state is certain to have one particular number of excitations, there is always a finite probability to have one more, or one less excitation than expected. The inset of Fig.~\ref{FigDminDmax}(b) shows the probability distribution for the locations of the excitations in the $5$-excitation ground state.

Clouds in a magneto-optical trap (MOT) are typically much larger in size, with diameters of $O$(1 mm). The $45 \mu$m excitation volume therefore needs to be created by shaping the intensity profiles of the excitation lasers. By passing the 780 nm laser through an aperture, it is possible to create a beam of width $45 \mu$m and approximately uniform intensity locally within the atom cloud. Intersecting this beam with a small, focused beam with diameter $4 \mu$m of the 480 nm laser near the center of the atom cloud results in the desired 1D excitation volume. Inside this volume the atomic density is practically uniform, and with typical densities of a MOT of $n = O(10^{10} - 10^{11}) \mathrm{cm}^{-3}$ there are approximately $10$ to $60$ atoms at uniformly distributed positions contained within the excitation volume. Figure \ref{FigDminDmax}(c) shows the critical detunings for such a system for the particular case of $30$ atoms. The probability distributions form peaks with distinctly asymmetric shapes, which are a result of the sharp boundaries of the excitation volume. The inset of Fig.~\ref{FigDminDmax}(c) shows the probability distribution for the locations of the excitations in the $5$-excitation ground state, which also exhibit the characteristic asymmetry due to the sharp boundary of the excitation volume.

As a final note to this section, we would like to mention that it is in principle possible to change system parameters such as the dimensions of the excitation volume. However, due to the $6$-th power scaling of the Van der Waals interaction energies this has a large effect on the other system parameters. For instance, changing all length scales by a factor $\lambda$ is equivalent to changing all time scales by a factor of $\lambda^6$, and consequently all frequencies such as $\Omega$ and $\Delta$ need to be scaled by a factor $\lambda^{-6}$. Even a simple change of the excitation volume length from $45 \mu$m to $90 \mu$m would therefore require all timescales to be multiplied by a factor of $64$ in order to obtain the same results. The chirps that we will construct for our particular excitation volume and particular experimental setup can therefore be said to be truly \textit{tailor made}.

\end{section}

\begin{section}{Performing the chirp}\label{SecPerformingChirp}
\begin{subsection}{Chirp properties}\label{SecChirpProperties}
Having outlined the properties of the physical systems of interest, we proceed by composing a realistic chirp of the laser detuning $\Delta$ and intensity $\Omega$. The basic ingredients for such a chirp are as follows.

\begin{enumerate}
\item{Start by switching on the laser and increase the coupling strength  $\Omega$ from zero up to some final value. This is done at a large, negative detuning with all particles starting in their atomic ground state, coinciding with the many body ground state.}
\item{Once the coupling $\Omega$ is sufficiently high, the detuning can be increased from its initial value towards its final target value $\Delta_f$ between $\Delta_{m-1}^m$ and $\Delta_m^{m+1}$.}
\item{Switch off the laser, by lowering the coupling strength back to $\Omega = 0$ again.}
\end{enumerate}
The above three steps should be carried out as adiabatically as possible, yielding a crystal state at the end of the chirp, such as depicted in the insets of Fig. \ref{FigDminDmax}. In Figure \ref{FigSimResults1}, panels (a), we show various typical chirps, modified by a computer algorithm to maximise adiabaticity given a constrained time frame of $\sim5 \mu s$ in which it is to be carried out. Each chirp is tailor made for a specific system. The remaining panels (b) and (c) of Fig. \ref{FigSimResults1} will be discussed in Sec. \ref{SecResults}.

Starting with step \textit{(i)} of the chirp, the first feature to be noted is the shape of $\Omega(t)$ in the first section, which is roughly exponential. As the coupling is still small, the energy levels are closely spaced together, and only small changes to the Hamiltonian can be made without making transitions to excited states. When the coupling is stronger, the ground state gets pushed to lower energies and separates further from the other energies, and as a consequence we can make larger changes to the Hamiltonian and increase the rate of change of the coupling. As the final value is reached, the ramp abruptly stops and $\Omega$ remains constant. Since the Schr\"odinger equation is a first order differential equation in time, such sharp changes in slope do not affect adiabaticity and there is no need to smoothen the chirp.

When in the second step the detuning is increased, we found that the maximum slope varies non-monotonically throughout the chirp. Regions where the detuning can only be changed relatively slowly are not necessarily correlated to the critical detunings $\Delta_{n}^{n+1}$ described in Sec. \ref{SecPhysicalSystems}. This is a sign of a complicated underlying energy landscape, whose details are beyond the scope of this paper.

The most critical step of the chirp is the final step, switching off the laser and lowering $\Omega$ back to zero again. When the detuning is large and positive, the energy scales in the system become larger, such that the coupling strength becomes relatively smaller. Moreover, in the case of a non-uniform density, such as the ODT or BEC, the $\sqrt{M_k}$ enhancement of the coupling in Eq. (\ref{SuperatomOmega}) necessitates a very fine control over the final stages of switching off the coupling, as even small intensities can still couple strongly to the central, high density region of the excitation volume. Because of these two phenomena, step $(iii)$ takes up by far the greatest amount of time of the total crystal creation procedure. Typically, the time scale for switching off the laser is an order of magnitude larger than the combined time for switching it on and chirping the detuning. Again it should be noted that the rate of change in $\Omega$ varies, as another sign of a complicated underlying energy landscape. We also found that it is favourable to \textit{decrease} the detuning slightly while ramping down the intensity.


\end{subsection} 

\begin{subsection}{Simulating excitation dynamics}
We now have all the necessary ingredients to perform a realistic chirp and simulate an attempt to create a Rydberg crystal in each of the four systems described in Sec. \ref{SecPhysicalSystems}: an optical lattice, a BEC and thermal clouds in an ODT and a MOT. We aim to create 5 or 6 excitations in total, due to limitations imposed by the computational complexity.

Following the ideas outlined in Sec. \ref{Systemdescription}, the simulation can be broken down into two parts. The first part consists of simulating the time evolution of the excitation dynamics of a given set of atoms at fixed positions. Then, in the second part a Monte Carlo integration over particle distributions is performed, summing many simulations for different atomic positions weighted by the probability that a certain configuration occurs.

For the simulation of the excitation dynamics for a fixed configuration, we perform a simulation of the full quantum many particle state, governed by the Schr\"odinger equation with Hamiltonian (\ref{EqH}). We use a fifth order adaptive stepsize Runge Kutta time integration \cite{Press92}, with the Cash-Karp parameter scheme \cite{Cash90}. For the optical lattice we assume $45$ particles in the excitation volume, whereas in the case of the BEC and the ODT we assume $N = \mathcal{O}(10^3 - 10^4)$ atoms in the excitation volume, and $30$ in that of the MOT. Since the parameter space for $N$ two level systems scales as $2^N$, we need to make some approximations to make the simulations tractable. The first approximation consists of reducing the $\mathcal{O}(10^3 - 10^4)$ atoms in the ODT and BEC to $40$ superatoms, by employing the algorithm suggested in Ref. \cite{Robicheaux05} where we recursively replace the closest spaced pair of (super)atoms by a weighted average position. In the end, we replace the couplings to the superatoms by the $\sqrt{M_k} \Omega$ coupling of Eq. (\ref{SuperatomOmega}), where $M_k$ denotes the number of atoms represented by the $k$-th superatom. But even the resulting $2^{40}$ states space is too large to handle, and we reduce the number of states further following the suggestions put forth in Refs. \cite{Weimer08, Wuster10, Pohl10}, where we omit all states with more than a certain maximum number of excitations ($>8$ in our case), and also omit all states where the excitations are spaced too close together. This latter approximation is made under the assumption that these states will never be significantly populated due to the Rydberg blockade. These three approximations reduced the state space to $O(10^{5})$ states, a number low enough to allow for a few hundred simulations per day on an $8$ core system.

\end{subsection} 

\begin{subsection}{Results}\label{SecResults}

For each of the systems described in Sec. \ref{SecPhysicalSystems} we have performed simulations of a tailor made chirp of laser detuning and intensity. The laser chirp is pre-determined per system by an algorithm that attempts to maximize adiabaticity, while keeping constraints on the total duration of the chirp. In this way, we have constructed chirps that create $6$ excitations in the optical lattice and BEC, and $5$ excitations in the thermal clouds in an ODT or MOT. All chirps have a duration of about $5$ $\mu$s.

For the optical lattice, the simulation only needed to be performed for a single configuration of atoms, since we assumed that the ground state atoms are at very well localised positions. For the other three systems we have performed many Monte Carlo simulations to account for the probabilistic distribution of the ground state particles. At the beginning of each simulation, we draw new particle positions from the appropriate particle distribution, and then simulate the time dynamics when this particular atom configuration is subjected to the laser chirp. In total, we have performed $600$ Monte Carlo samplings for the BEC, $1650$ samplings for the ODT and $3300$ samplings for the MOT.

After all the simulations are completed, the results can be aggregated in order to compute expectation values of various observables. In particular, for each system we compute expectation values for the number of excitations, the excitation density, and the pair correlation function.

These expectation values can be constructed using the excitation number operator $\exnr{r}$, which records the number of excitations measured at the (1D) distance $r$ from the origin (not to be confused with the many particle position vector $\rvec$). We define this operator through its matrix elements in the basis (\ref{EqBasis}):
\begin{equation}\label{EqExnr}
\bra{\rvec' \otimes \nvec'} \exnr{r} \ket{\rvec \otimes \nvec} = \delta(\rvec - \rvec') \delta_{\nvec \nvec'} \indfn(\rvec, \nvec),
\end{equation}
where $\indfn(\rvec, \nvec)$ is an \textit{indicator function}, which takes on the value of $1$ if the state $\ket{\rvec}\otimes\ket{\nvec}$ has an excited particle at position $r$, and $0$ otherwise. The corresponding expectation value is found from Eq. (\ref{EqObs}):
\begin{equation}\label{EqExcDens}
\langle \exnr{r} \rangle = \int \drvec \ \sum_{n=0}^{2^N - 1}|c_0(\rvec)|^2 \ |c_{n, \rvec}(t)|^2 \ \indfn(\rvec, \nvec).
\end{equation}
Note that the coefficients $c_0(\rvec)$ follow from the Monte Carlo sampling, whereas the coefficients $c_{n, \rvec}(t)$ are calculated in the simulation of the time dynamics. 

Using the operator $\exnr{r}$ we can additionally compute the second-order density correlation function \cite{Glauber}
\begin{equation}\label{EqG2rr}
G^{(2)}(r, r') = \langle \exnr{r} \exnr{r'} \rangle - \delta(r - r') \langle \exnr{r} \rangle,
\end{equation}
which measures the conditional probability of finding an excitation at point $r'$, given that there already is an excitation found at $r$. It is more convenient to compute the integrated second-order density correlation function
\begin{equation}\label{EqG2}
G^{(2)}(r) = \int G^{(2)}(r, r + R)\  \mathrm{d}R,
\end{equation}
which measures just the probability of finding an excitation a distance $r$ away from any other excitation. Calculated on a discrete grid, the correlation function $G^{(2)}$ can be thought of as a histogram of pair separations.

The final observable we calculate is the probability $p(m)$ of finding $m$ excitations, which can be calculated as
\begin{equation}\label{EqOccnr}
p(m) = \int \drvec |c_0(\rvec)|^2 \sum_{n=0}^{2^N-1}|c_{n, \rvec}(t)|^2 \ \indfm(\nvec),
\end{equation}
where we have defined an additional indicator function $\indfm(\nvec)$, which takes on the value $1$ if the state $\nvec$ has $m$ excitations, and $0$ otherwise.

\begin{figure}[h!]
\begin{center}
\includegraphics[width=0.95 \textwidth]{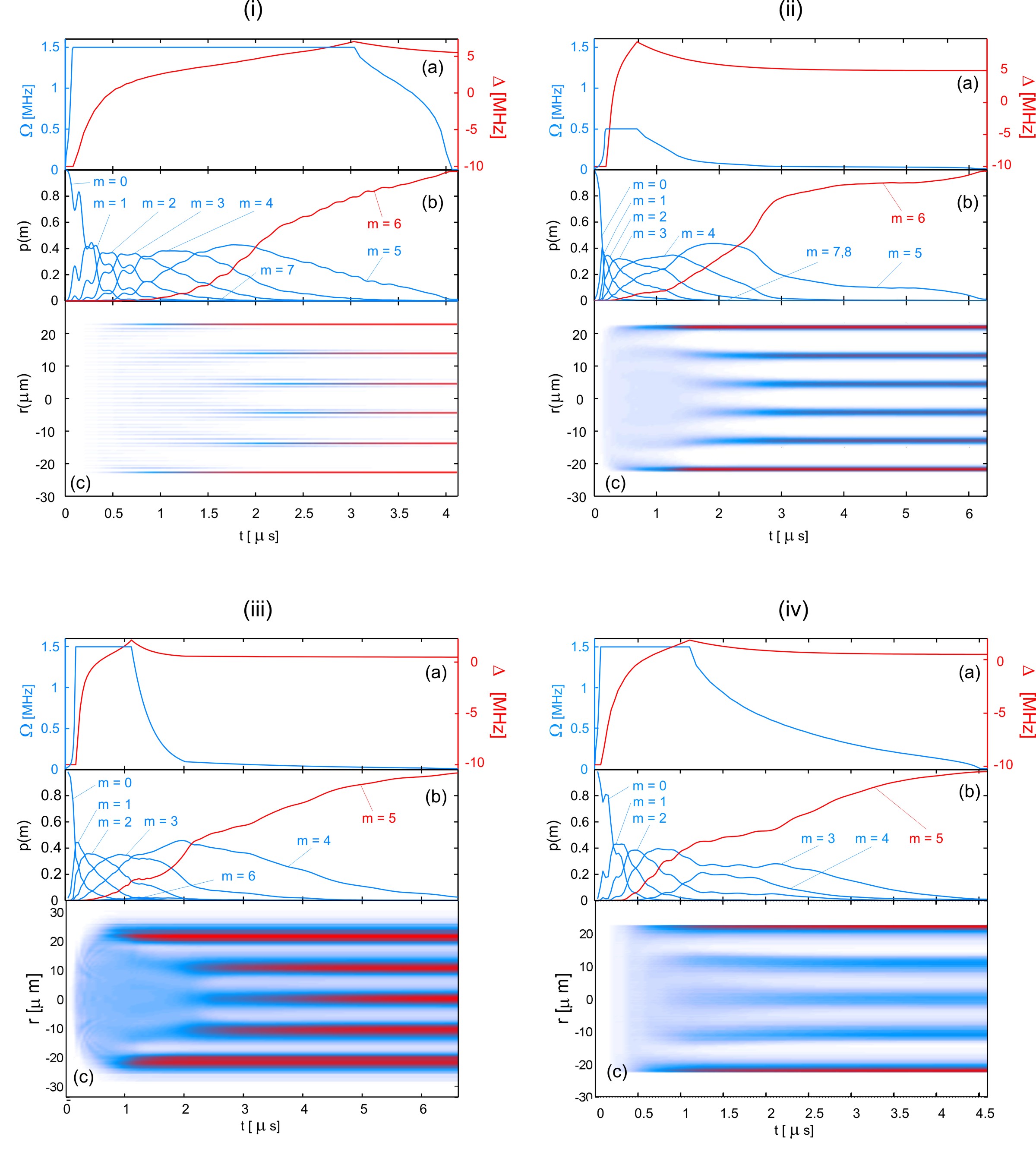}
\caption{Simulation results for Rydberg crystal creation in each of the four systems discussed in the text: (i) optical lattice, (ii) Bose-Einstein condensate, (iii) thermal cloud in an ODT, and (iv) thermal cloud in a MOT.
All results are shown as a function of time. In each subfigure, the top panel, labeled (a) shows the laser intensity $\Omega$ and detuning $\Delta$, the center panel, labeled (b) shows the probability $p(m)$ of finding $m$ excitations. The bottom panel, (c), shows the spatial excitation density $\langle\exnr{r}\rangle$ during the chirp, indicated with colors ranging from white (lightest) to blue to red (darkest), where darker colors indicate a higher probability of finding an excitation at that position and time.} \label{FigSimResults1}
\end{center}
\end{figure}

We are now ready to discuss the results of the simulations. Figure \ref{FigSimResults1} shows the calculated expectation values for each of the four systems we are interested in: \textit{(i)} optical lattice, \textit{(ii)} Bose-Einstein condensate, \textit{(iii)} thermal cloud in an ODT, \textit{(iv)} thermal cloud in a MOT. For each system, the top panel, (a), shows the chirp of the laser intensity $\Omega$ and detuning $\Delta$, the second panel, labeled (b), shows the expectation values $p(m)$ of Eq. (\ref{EqOccnr}) of the number of excitations, the third panel, (c), shows the excitation density (\ref{EqExcDens}), where all expectation values are plotted as a function of time.

\begin{figure}[h]
\begin{center}
\includegraphics[width=8cm]{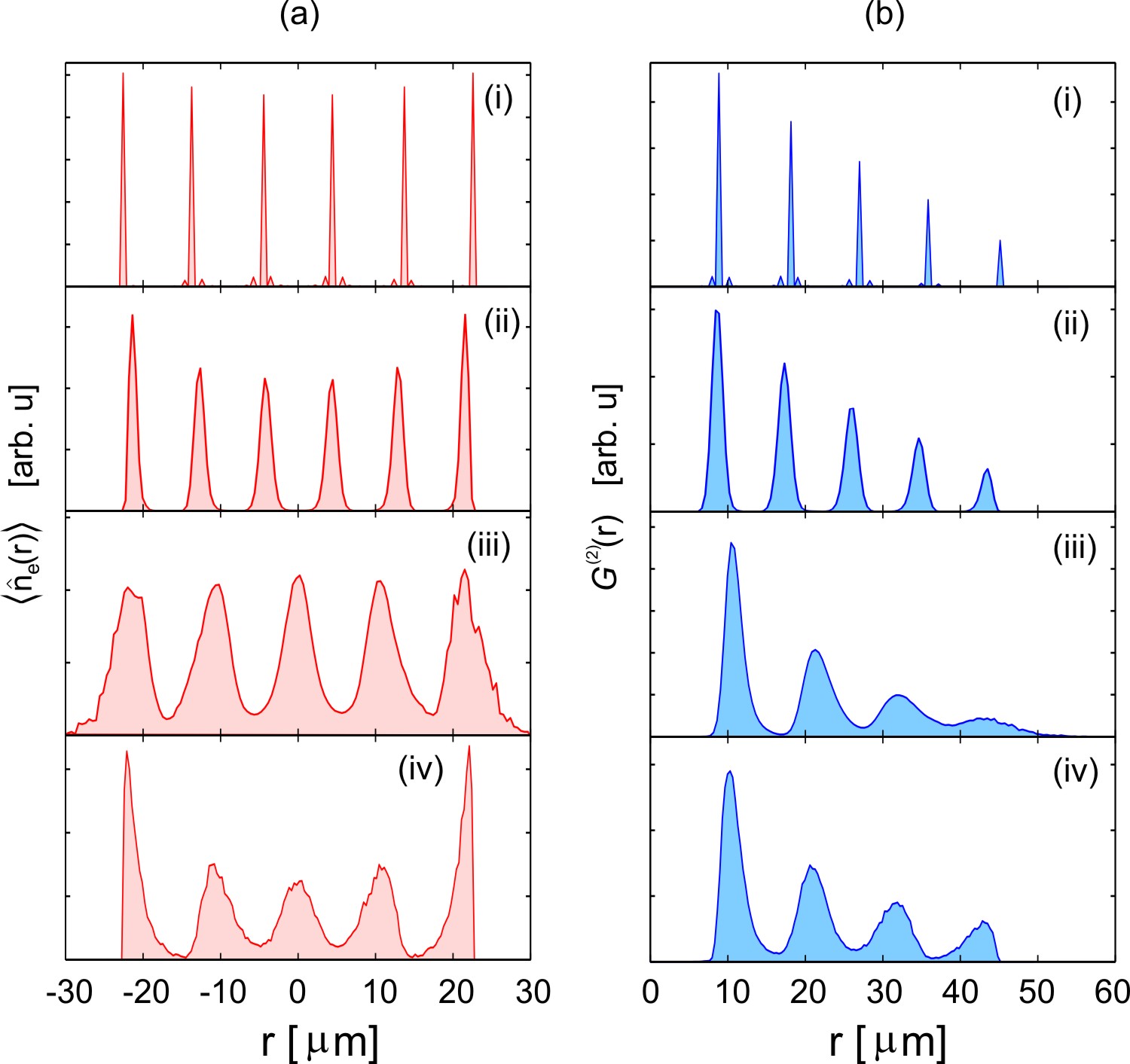}
\caption{(a): Excitation density, and (b): correlation function $G^{(2)}(r)$ at the end of the chirp for each of the four systems under consideration: (i) optical lattice, (ii) Bose-Einstein condensate, (iii) thermal cloud in an ODT and (iv) thermal cloud in a MOT. Note that the optical lattice and BEC have $6$ excitations, whereas the ODT and MOT have $5$ excitations.} \label{FigSimResults2}
\end{center}
\end{figure}

There are a number of similarities and differences to be noted between the four systems. Starting with the top panel (a), we see that the shape of each chirp follows roughly the three-step scheme described in Sec. \ref{SecChirpProperties}, although the length of each segment varies between systems. Typically, the third step of switching the laser off takes the longest time.

Panels (b) and (c) of Fig. \ref{FigSimResults1} show clearly what happens during the three steps of the chirp. When the laser is switched on in the first step, the initial state $\ket{\Ovec}$ with all atoms in the ground state becomes a superposition of states with $0$ or $1$ excitations. Then, as the detuning is increased while the intensity is at its maximum, states with higher numbers of excitations start to mix into the system state. No particular number of excitations has a high probability of occurring (panel (b)), and the location of the excitations is also spread out across the excitation volume (panel (c)). Only when in the final step the laser is switched off do the excitations localise at regularly spaced positions, and the Rydberg crystal takes shape. First, the outermost excitations localise, and then excitations crystallise progressively inwards towards the center of the excitation volume. In panel (i.c) for the excitation density of the optical lattice, one can distinguish the individual atoms at each lattice site, and see how the excitations are initially shared among all atoms and crystallise only at the end of the chirp.

The middle section of the chirp takes the shortest time in the systems with high particle numbers, the BEC (ii) and the ODT (iii), as a result of the enhanced $\sqrt{M_k}$ coupling in the superatom picture [see Eq. (\ref{SuperatomOmega})]. The effect of this enhanced coupling is also clearly visible in panels (a) of Fig. \ref{FigSimResults1} for systems (ii) and (iii), where a disproportionally long time is needed to ramp down the last few percents of laser intensity. The center panels (b) show that in this long flat tail of the intensity chirp there are still large changes in the excitation number probabilities.

Fig. \ref{FigSimResults2} shows the excitation density (\ref{EqExcDens}) at the end of the chirp (column (a)), and correlation function (\ref{EqG2}) at the end of the chirp (column (b)), for each of the four systems (i) - (iv) of interest.

Starting with the excitation density in the optical lattice in the top left panel (a.i), we see that the excitations are more or less perfectly localised on individual lattice sites. Upon close inspection there are small probabilities discernable, for finding the excitation at neighbouring lattice sites, indicating that the time evolution has not been perfectly adiabatic.
The excitation density of the BEC in panel (a.ii) also exhibits distinct and well-separated peaks. However, the excitation density of the thermal cloud in an ODT in panel (a.iii) shows peaks that, although clearly visible, are not entirely separated spatially. This is due to the boundaries of the excitation volume not being as sharply defined as those of the BEC. In the ODT, the spread in the position of the outermost particles is largest of all four systems. Even the MOT excitation density in panel (a.iv) shows better separated peaks. Due to the sharp boundary of the excitation volume in this system (see Sec. \ref{SecPhysicalSystems}) the peaks show a roughly similar shape to those of the BEC. However, they exhibit a larger spread around their mean positions, reflecting the much lower number ($30$) of particles present in the excitation volume. In effect, the low particle number offers fewer positions for the excitations to localise on, and the optimal positions leading to the lowest energies are often not occupied in a particular Monte Carlo simulation.

Finally, moving to column (b) of Fig. \ref{FigSimResults2}, showing the correlation function for each system, we can extract the following information. Interpreting the correlation functions $G^{(2)}$ as histograms for the distances between pairs of excitations, we clearly see that there are never two excitations within one peak of the excitation densities in column (a). Should this have been the case, then we would find nonzero probabilities at distances smaller than the inter-peak distance of $\sim 10\mu m$.

The first peak in $G^{(2)}$ appearing around $10 \mu m$ is the distribution of pair separations of excitations in \textit{neighbouring} peaks in the excitation density. Likewise the second peak in $G^{(2)}$ is the distribution of pairs in peaks separated by a single other peak, followed by the pair distributions of excitations in peaks separated by two peaks, etc. The height of the peaks in the correlation function $G^{(2)}$ directly reflects the number of pairs of peaks separated by $0, 1, 2, ..$ other peaks, leading to a  linear decay in height. The width of the peaks signifies how regularly the excitations are spaced. Again, in case (iii) of the ODT the peaks are the least sharply defined, as a direct result of the excitation volume not being bounded sharply. In case (i), the optical lattice, there are small sub-peaks visible again, which are correlations between the large peaks and sub-peaks of the excitation density in panel (a.i).

\end{subsection}

\end{section} 

\begin{section}{Higher dimensional systems}\label{SecHigherDimensionalSystems}

\begin{figure}[h]
\begin{center}
\includegraphics[width=8cm]{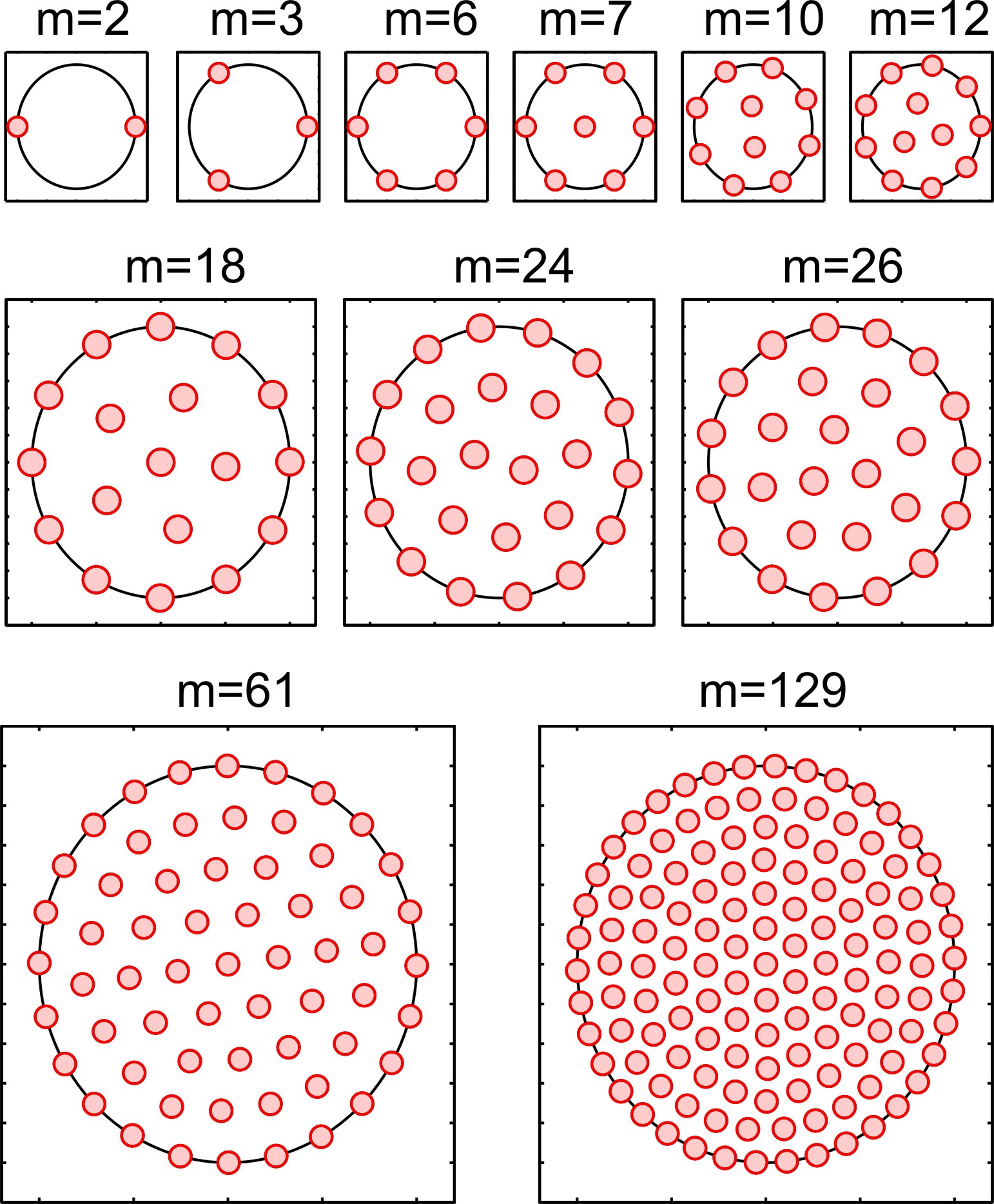}
\caption{Examples of 2D crystal structures as function of Rydberg atom number $m$. Initially, when there are few Rydberg atoms, all excitations reside on the boundary of the excitation volume in order to maximise the interparticle distance. As the number of Rydberg atoms confined in the circular volume increases, the inner region of the excitation volume starts to fill as well. At $m=7$ the first particle appears in the center, followed at $m=10$ and $m=12$ by a second and third one respectively. At $m=18$, an additional shell starts to form at the center, progressively filling up ($m =24, 26$,..) until there are so many particles in the interior that boundary effects become negligible and a transition to bulk behaviour is observed, where a hexagonal lattice is formed in the central region ($m = 61$ and $m=129$). The corresponding energies of these configurations are shown in Fig.~\ref{fit2D3D}(a).} \label{Fig2D}
\end{center}
\end{figure}

In the previous section, we focused in detail on the adiabatic creation of one-dimensional Rydberg crystals. However, the dimensionality of the crystal structures can be controlled by the shape of the volume set up by the excitation lasers. As 1D structures are produced by a needle-shaped light field, so can 2D structures be created by a sheet of light intersecting the atomic cloud. 3D structures can arise in a light volume where excitations are not prohibited by dimensional restrictions. In this section we take a closer look at what type of 2D and 3D structures may arise when applying the same type of laser excitation techniques in 2D and 3D light volumes, and investigate the energy of the ground states as function of the number of Rydberg atoms. This latter quantity is of importance when determining the critical detunings $\Delta_m^{m+1}$ described in Sec \ref{SecPhysicalSystems} at which the system transitions from an $m$-excitation ground state to an $m+1$-excitation ground state.

\begin{figure}[h]
\begin{center}
\includegraphics[width=12cm]{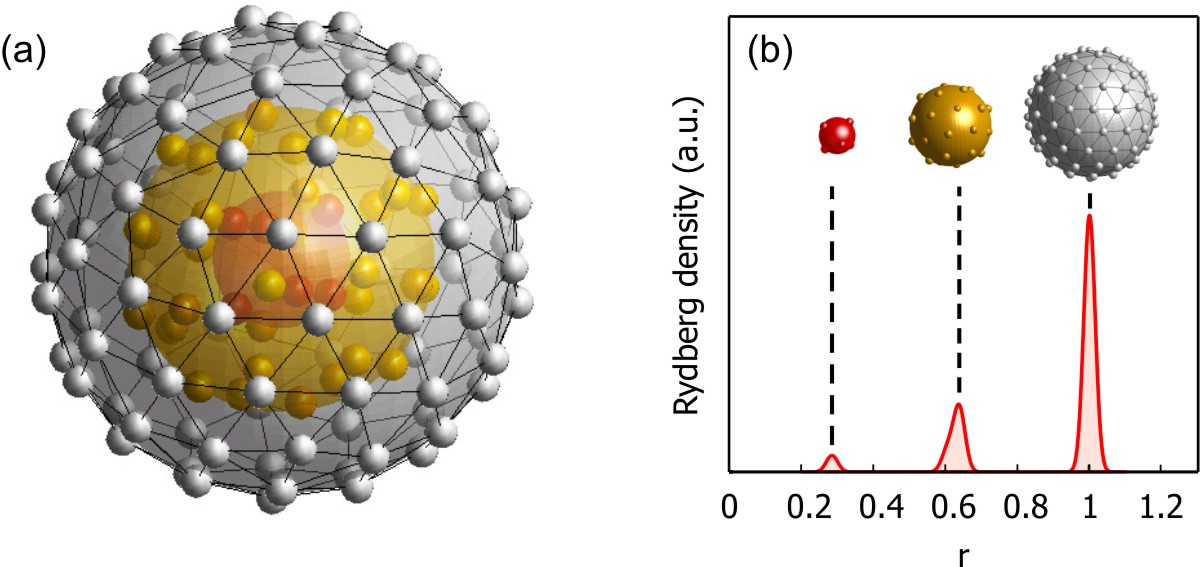}
\caption{Example of a 3D crystal structure. (a) displays the structure as a whole, while in (b) the inner shells can be seen, with an indication of the average distance of the shell to the center of the structure.} \label{Fig3D}
\end{center}
\end{figure}

While in the 1D case the crystal structures are basically similar to regularly spaced chains of Rydberg excitations, in higher dimensions the structures can be quite complicated. However, for simplicity, we assume that the crystal structures with $m$ excitations are confined by a circular plane (2D) or spherical volume (3D), which have hard wall boundaries and where the interior can be considered as homogeneous with no external potential.


We calculate the optimal crystal structures by using a Conjugate Gradients \cite{Press92} minimization of the energy of $m$ point particles interacting via a repulsive two-body Van der Waals $C_6 / r^6$ potential:
\begin{equation}
E_{int}(\rvec_1, \rvec_2, .. , \rvec_m) = \mathop{\sum_{i=1}^m}_{j > i} \frac{C_6}{|\rvec_i - \rvec_j|^6}.
\end{equation}
The minimization is performed with respect to the particle positions $\rvec_i$. Since this energy represents a very complex $2m$ (2D) or $3m$ (3D) dimensional  energy landscape, with many local minima, we have performed hundreds of minimizations for each Rydberg particle number $m$ with random initial conditions, and recorded the lowest energy configuration thus found as a good approximation for the global minimum. There exists a large body of literature in the field of Coulomb crystals where similar procedures are found to find crystal structures of trapped ions, see e.g. Refs. \cite{Totsuji02, Pohl04, Ludwig05, Mortensen06, Apolinario07}, and references therein.

\begin{figure}
\begin{center}
\includegraphics[width=\columnwidth]{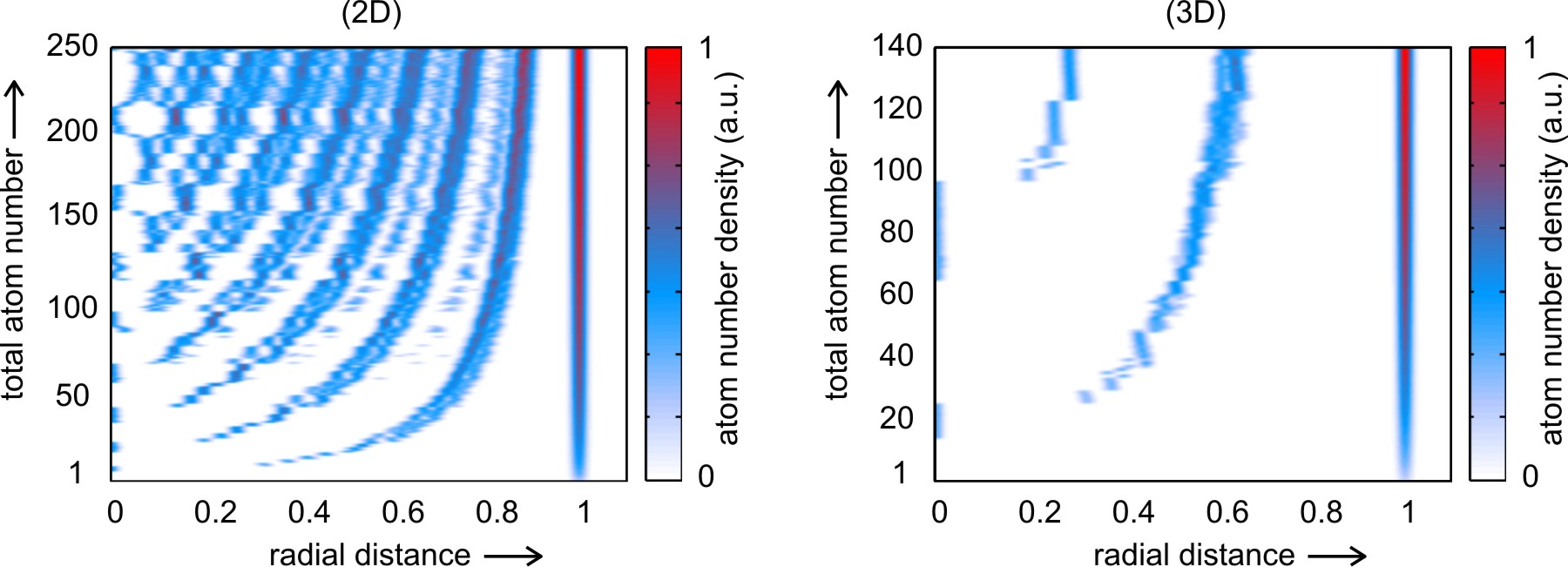}
\caption{Shell formation for a 2D (left) and 3D (right) Rydberg lattice. The Rydberg atom number density is indicated as a function of the radial distance to the center of the system (horizontal axis) and for a varying total number of Rydberg atoms (vertical axis). Each horizontal slice of the figure corresponds to particular configuration, as indicated for instance in Figs.~\ref{Fig2D} and \ref{Fig3D}. When the number of atoms is increased, additional shells appear (blue / red fuzzy lines) that move toward greater radial distance to make room for yet more inner shells. A large part of the atoms resides on the boundary of the system, corresponding to the bright line at radial distance $r = 1$. In the 2D system, the onset of lattice formation and breakdown of shell structure can be observed near $r = 0$ for high atom numbers where the shell structure becomes more diffuse.}\label{fi:shell_formation}
\end{center}
\end{figure}

In Figs.~\ref{Fig2D} and \ref{Fig3D} examples can be seen of crystal structures of Rydberg excitations, for a given number of $m$. Typically, we observe in $2D$ that at low particle numbers $m$ all particles are located on the boundary of the volume, maximizing the distances between the nearest neighbours. As $m$ is increased the boundary fills up, until at a certain point it becomes energetically favourable for the new particle to be placed in the center of the volume. As the particle number is increased further, more particles will be located near the center, forming a ring shaped structure. This process continues until, in turn, a particle appears in the center of this ring again, starting the cycle anew. For even larger particle numbers, there are so many particles located in the interior that boundary effects become negligible and the particles assume a hexagonal lattice structure typical of bulk behaviour.

Generally speaking, for larger numbers of excited atoms a competition occurs between lattice structures in the interior, and shell structures near the boundaries of the system, an effect familiar from Coulomb crystal studies \cite{Totsuji02, Mortensen06}. Positions on the boundary of the system are energetically favourable due to the (partial) absence of neighboring atoms, leading to boundary effects (shell structures), propagating inward until the boundary is sufficiently far away such that bulk behavior (lattice structures) start to dominate.  For a sufficiently large number of Rydberg atoms, the structure can then be regarded as a (quasi)  periodic, homogeneous lattice. However, for small atom numbers we enter the regime of mesoscopic physics, where we create symmetric few-body clusters.

In $3D$ systems we see the same chain of events, with consecutive shells  forming when the particle number is increased. Here, we expect lattice formation in the interior only at much higher Rydberg atom numbers since surfaces of spheres can accomodate much more particles than the circular boundary in the $2D$ case. We have verified that up to $m = 1000$ we still see distinct shell structures in $3D$.

As a function of the number of Rydberg atoms in the lattice, we show in Fig.~\ref{fi:shell_formation} where the atoms will be located as a function of the radial distance to the center of the excitation volume. Both in $2D$ and $3D$ the formation of shells is visible for increasing Rydberg atom number $m$.

\begin{figure}[h]
\begin{center}
\includegraphics[width=7cm]{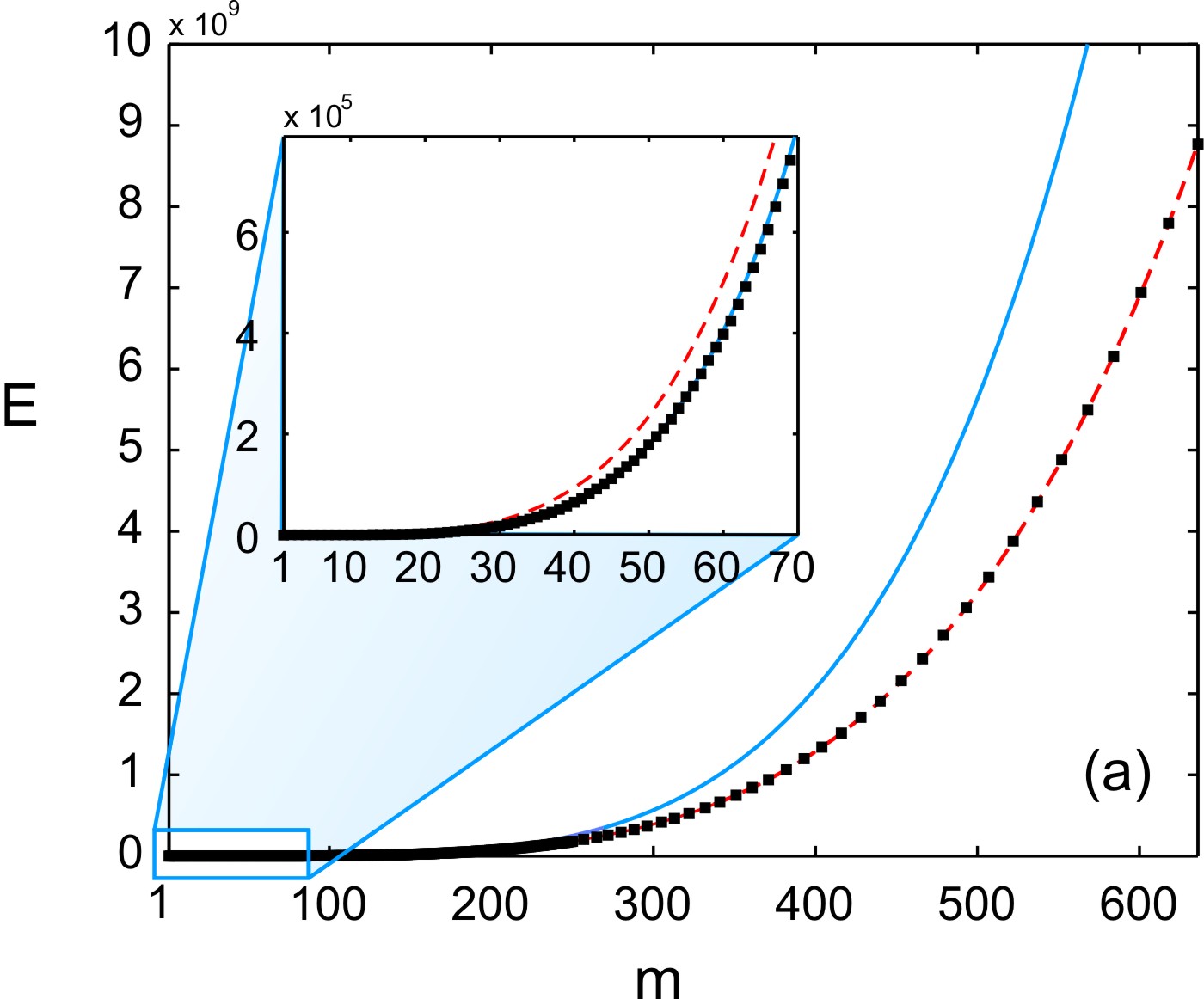}
\includegraphics[width=7cm]{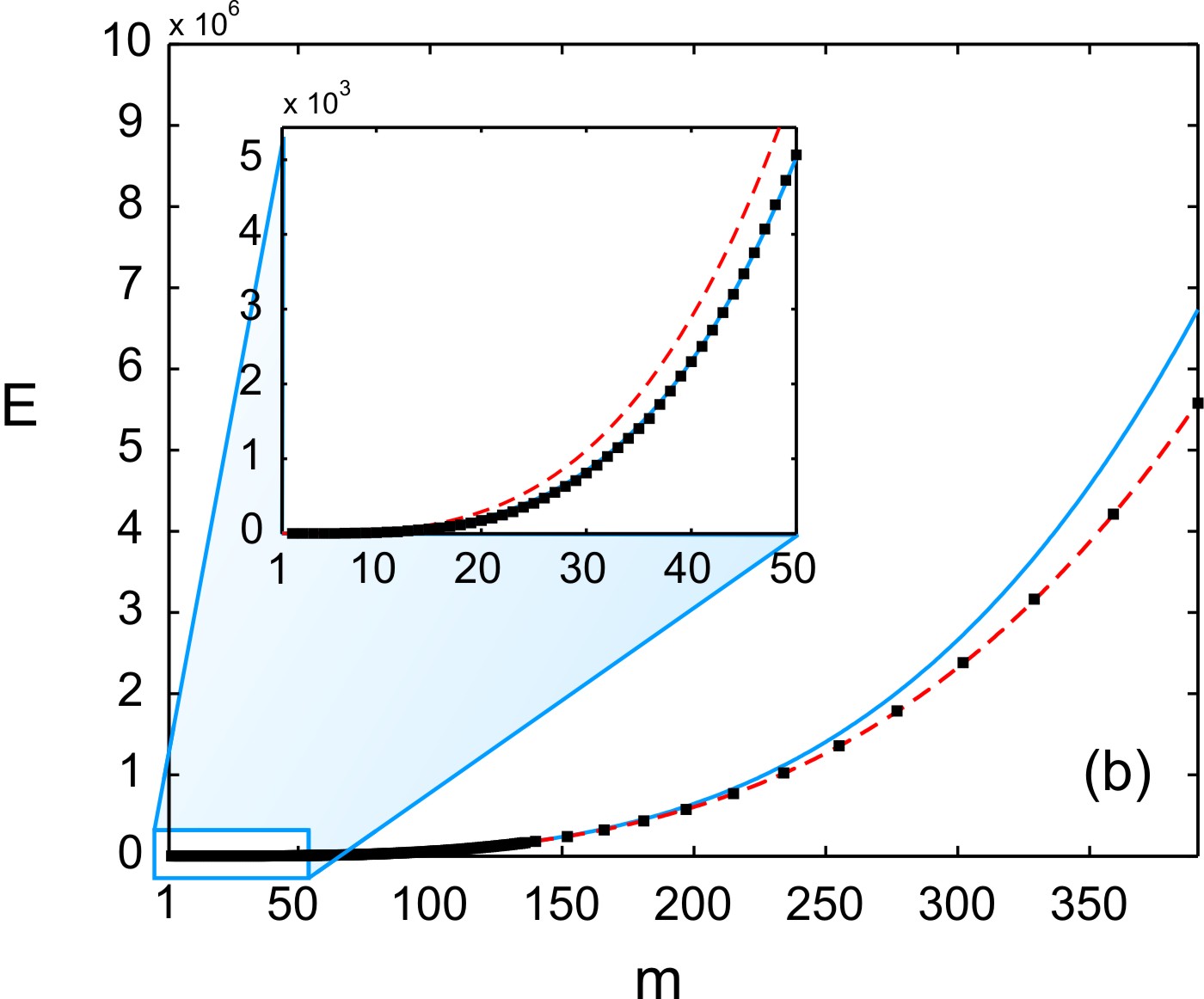}
\caption{Minimum energy $E$ as function of the number of Rydberg atoms $m$, for the distribution of $m$ atoms in (a) a two-dimensional circular plane, (b) a three-dimensional spherical volume. The insets display the behavior for smaller atom numbers. The continuous lines represent fits according to the mesoscopic (blue, solid line) and bulk (red, dashed line) scaling laws.} \label{fit2D3D}
\end{center}
\end{figure}

In Fig.~\ref{fit2D3D} the minimum energy $E(m)$ is shown as a function of the particle number $m$, for the 2D and 3D case. The energy solutions show three distinct regimes: the few-particle regime where boundary effects dominate, the bulk regime with many particles, and inbetween the mesoscopic regime which connects the two limits. Each regime has a different scaling of energy versus particle number. In the bulk regime the energy scales as $m^4$ in 2D and as $m^3$ in 3D. This scaling behavior assumes a homogeneous distribution of the particles with a mean interparticle separation, and where each particle has a fixed number of neighbours. Furthermore, we assume only nearest neighbour interactions. In the few-particle regime, edge effects dominate, which results in a relatively larger interparticle separation as in the bulk regime. However, despite a lower than bulk energy, the energy increases more rapidly with $m$. In the mesoscopic regime, the system is inbetween the edge-dominated and bulk dominated regimes, which results in energy scaling laws as $\sqrt m m^4$ for 2D and $\sqrt m m^3$ for 3D.

\end{section} 

\begin{section}{Conclusions and outlook}\label{SecConclusions}
While crystalline phases have been predicted for strongly interacting Rydberg gases, it is not straightforward to achieve such an ordered structure starting from a disordered gas. A practical scheme to enter this regime has been proposed by Pohl {\it et al.}~\cite{Pohl10}, using chirped laser pulses.
In this paper, we investigate the feasibility to create one-dimensional Rydberg crystals in four distinct cold atomic systems, which vary in density and temperature, and demonstrate the importance of an excitation scheme which is exactly tailored to the specific system parameters. Even for a system that is relatively low in density and high in temperature, such as a magneto-optical trap, it is possible to design an excitation scheme that creates spatially ordered structures. A successful experimental realization of one-dimensional Rydberg crystals allows for a direct comparison to predicted analytical properties in a lattice~\cite{Lesanovsky11}. For crystalline Rydberg structures in two or three dimensions, we calculate some basic properties such as the spatial structure, the corresponding ground state energy, and shell structure formation as function of the Rydberg atom number.
\end{section} 

\section*{Acknowledgements}
We thank Arieh Tal for assisting us with computation on the Polyxena cluster, generously provided by the group TPS of the Applied Physics department. RMWvB and SJJMFK acknowledge financial support from the TU/e Fund for Excellence.

\appendix

\begin{section}{A closer look at the frozen gas approximation} \label{dephasing}

Throughout this paper, we assumed that the system is sufficiently cold such that we can make the frozen gas approximation, and hence neglect the kinetic energies of the atoms during the course of a typical experiment lasting a few microseconds. In a typical experiment, the laser detuning is swept through a few MHz within a few $\mu$s, and in this Appendix we investigate whether thermal motion under MOT conditions can significantly affect the final state after a sweep.

\begin{figure}[h]
\begin{center}
\includegraphics[width=14cm]{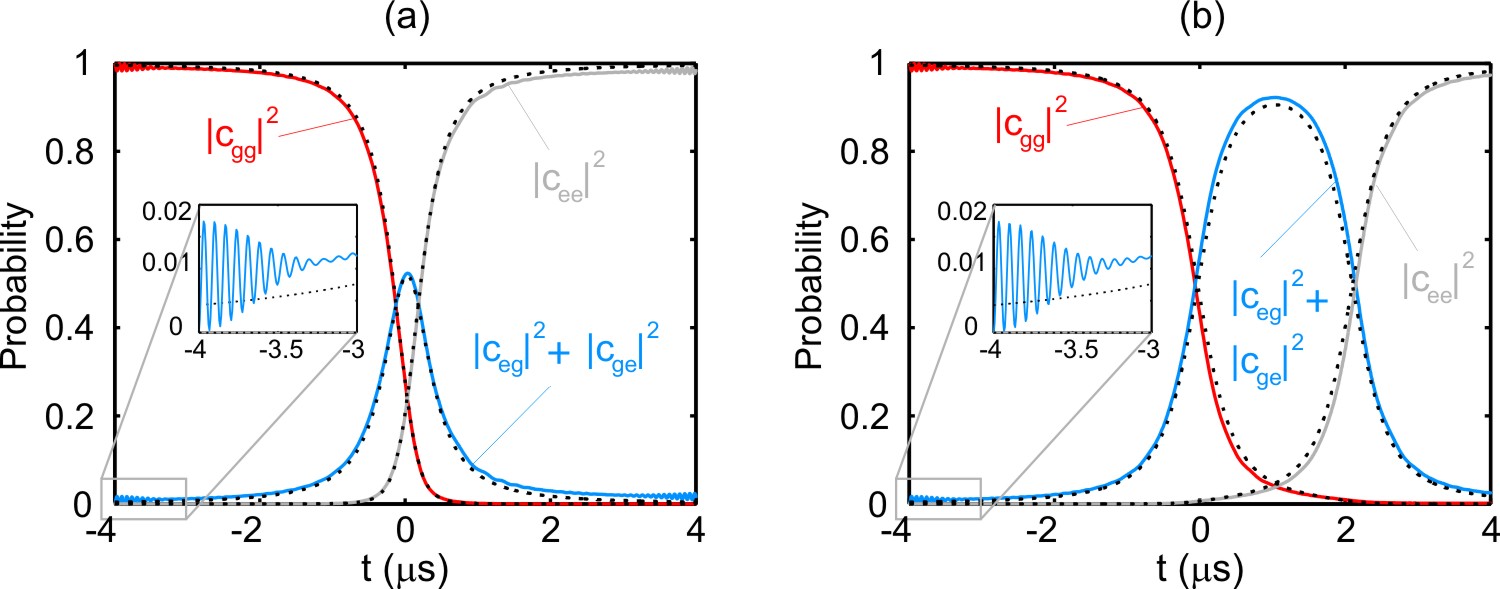}
\caption{Velocity-averaged two-particle squared amplitudes as function of time, during the course of an adiabatic chirped transition from the ground state to the excited Rydberg state. Shown are $|c_{gg}|^2$ (red), $|c_{ge}|^2+c_{eg}|^2$ (blue), and $c_{ee}|^2$ (gray). From $t=-4\, \mu \mathrm{s}$ to $t=4\, \mu \mathrm{s}$, $\delta$ is chirped from -15 MHz to 15 MHz in a linear fashion. $\Omega$ is held constant at 1.4 MHz. The interaction energy $V_{int}$ of the $c_{ee}$ state is (a) 0.2 MHz and (b) 7.8 MHz, corresponding to a distance of about 11 and 6 $\mu \mathrm{m}$ respectively. The absolute squares of the amplitudes are averaged over a Maxwell-Boltzmann distribution with a temperature corresponding to a Doppler limit of 0.5 MHz. Also shown are the instantaneous eigenstates of the ideal system with $v_{1}=v_{2}=0$ (black line). The decaying Rabi oscillations can be seen, which are found to damp out in a Gaussian fashion with a time constant $\tau \approx 0.45 \,\mu \mathrm{s}$, in agreement with Eq.~\ref{eq:dampedrabi}. Note that the oscillations in (a) experience an adiabatic revival at $t=4\,\mu \mathrm{s}$ because of the symmetry of the sweep.}
\label{fig:sweepdelta2particles}
\end{center}
\end{figure}

We consider a two particle two-level system with a time-dependent Rabi frequency and detuning, described by the coupled set of optical Bloch equations for the coupled two-particle amplitudes  $|c_{gg}|^2$, $|c_{ge}|^2$, $|c_{eg}|^2$, and $|c_{ee}|^2$, where subscripts $g$ and $e$ denote ground or excited states. The time-dependent equations are given by

\begin{eqnarray}
\label{eq:cggt_doppler}
i \frac{\mathrm{d}c_{gg}(t)}{\mathrm{d}t} &=& \frac{\Omega(t)}{2}(c_{ge}(t)+c_{eg}(t)), \\
\label{eq:cget_doppler}
i \frac{\mathrm{d}c_{ge}(t)}{\mathrm{d}t} &=& \frac{\Omega(t)}{2}(c_{gg}(t)+c_{ee}(t))+ (-\Delta(t)+k v_{2})c_{ge}(t), \\
\label{eq:cegt_doppler}
i \frac{\mathrm{d}c_{eg}(t)}{\mathrm{d}t} &=& \frac{\Omega(t)}{2}(c_{gg}(t)+c_{ee}(t))+(-\Delta(t)+k v_{1})c_{eg}(t), \\
\label{eq:cegt_doppler}
i \frac{\mathrm{d}c_{ee}(t)}{\mathrm{d}t} &=& \frac{\Omega(t)}{2}(c_{ge}(t)+c_{eg}(t))+(-2\Delta(t)+k v_{1}+k v_{2}+V_{int})c_{ee}(t).
\end{eqnarray}

The Doppler shifts $k v_{i}$ are added to the detuning, which allow for different velocities for particle $i=1,2$. A conservative estimate for these shifts is of order $3\cdot 10^6$ rad/sec or 0.5 MHz, resulting from the effective transition wavelength of 300 nm, that gives rise to the laser wavenumber $k$. The dephasing that arises is a result of the distribution of velocities $v_{i}$ for an ensemble of atoms, which we assume to obey Maxwell-Boltzmann statistics, and this will damp Rabi oscillations. The time scale associated with this damping is

\begin{equation}
\label{eq:dampedrabi}
\tau = \sqrt{ \frac{2 m (\Omega^2+\Delta^2)}{\Delta^2 k^2 k_B T} },
\end{equation}
with $m$ the atomic mass and $k_B$ the Boltzmann constant. However, damping will not be a limiting factor for the experimental conditions which we consider in this paper, provided that the chirp is wide enough, as can be seen in  Fig.~\ref{fig:sweepdelta2particles}. Here an adiabatic transition is shown for two ground state atoms to their excited Rydberg states, by presenting the velocity averaged two-particle amplitudes. The calculation is done for two different interparticle separations, and the squared amplitudes $|c_{gg}|^2$, $|c_{ge}|^2$ etc. are averaged over Maxwell-Boltzmann distributions, corresponding to the Doppler temperature, for velocities $v_i$. The figure shows damped Rabi oscillations. Nevertheless, the amplitudes are able to follow adiabatically the instantaneous eigenstates if the chirp is slow enough, and wide enough, meaning that $|\Delta(t_{\mathrm{initial}})-k v|\gg \Omega(t_{\mathrm{initial}})$ and $|\Delta(t_{\mathrm{final}})-k v|\gg \Omega(t_{\mathrm{final}})$. In conclusion, since the Rydberg crystal creation relies on adiabatic processes, dephasing of Rabi oscillations does not inhibit the crystallization process.

\end{section}

\end{document}